\documentclass{sig-alternate}
\usepackage{url}

\begin{document}

\newtheorem{definition}{Definition}

\newcommand{\CDIRECT}{d_A}
\newcommand{\CFUT}{o_A}
\newcommand{\REV}{r_A}
\newcommand{\LENGTH}{t}
\newcommand{\CREMV}{v_A(i)}
\newcommand{\CSETUP}{u_A}
\newcommand{\DIFF}{\theta_A(i,t)}
\newcommand{\DIFFA}{\theta_A(a,t)}
\newcommand{\CEXEC}{e_A(t)}
\newcommand{\CEFF}{f_F(i)}
\newcommand{\CREMVF}{v_F(i)}
\newcommand{\CREMVX}{v_x(i)}
\newcommand{\ECONOMICS}{q_A}

\newcommand{\MREMV}{\bar{v}_A(\TYPE)}
\newcommand{\MREMVX}{\bar{v}_x(\TYPE)}
\newcommand{\MEXEC}{\bar{e}_A(t)}
\newcommand{\MEXECX}{\bar{e}_x(t_x)}
\newcommand{\MEFF}{\bar{f}_F(\TYPE)}
\newcommand{\MSETUP}{\bar{u}_A}
\newcommand{\MSETUPX}{\bar{u}_x}
\newcommand{\MREMVF}{\bar{v}_F(\TYPE)}

\newcommand{\TYPE}{\tau_i}
\newcommand{\TYPEJ}{\tau_j}
\newcommand{\PTYPE}{p_i(\TYPE)}
\newcommand{\CLASSES}{C}
\newcommand{\CLASS}{c}
\newcommand{\ECLASS}[1]{E_{p_i^c}(#1)}
\newcommand{\DDT}{de\-fect-de\-tec\-tion tech\-nique}
\newcommand{\DIFFT}{\theta_A(\TYPE,t)}

\newcommand{\PCLASS}{q_i(\CLASS)}

\newcommand{\EREMV}{E_\alpha(i,\CREMV)}
\newcommand{\EEFF}{E_\sigma(i,\CEFF)}
\newcommand{\EREMVF}{E_\alpha(\CLASS,\CREMVF)}
\newcommand{\EREMVX}{E_\alpha(i,\CREMVX)}

%
\conferenceinfo{ISESE'06,}{September 21--22, 2006, Rio de Janeiro, Brazil.}
\CopyrightYear{2006} 
\crdata{1-59593-218-6/06/0009}  

\title{A Literature Survey of the Quality Economics of 
Defect-Detection Techniques}
%
%

\numberofauthors{1}
%

\author{
%
\alignauthor Stefan Wagner\\
       \affaddr{Institut f\"ur Informatik}\\
       \affaddr{Technische Universit\"at M\"unchen}\\
       \affaddr{Boltzmannstr.\ 3, D-85748 Garching b.\ M\"unchen, Germany}\\
       \email{wagnerst@in.tum.de}
}

\maketitle

\begin{abstract}
Over the last decades, a considerable amount of empirical
knowledge about the efficiency of \DDT s has been accumulated. Also
a few surveys have summarised those studies with different focuses,
usually for a specific type of technique. This work reviews
the results of empirical studies and associates them with a
model of software quality economics. This allows a better comparison
of the different techniques and supports the application of the model
in practice as several parameters can be approximated with typical
average values. The main contributions are the provision of average
values of several interesting quantities w.r.t.\ defect detection and
the identification of areas that need further research because of
the limited knowledge available.
\end{abstract}

\category{D.2.8}{Software Engineering}{Metrics}
\category{D.2.5}{Software Engineering}{Testing and Debugging}

\terms{Economics, Verification, Reliability}

\keywords{Software quality economics, quality cost, cost/benefit, 
          defect-detection
          techniques, literature survey}

\section{Introduction}
\label{sec:intro}

The economics of software quality assurance (SQA) are a highly
relevant topic in practice. Many estimates assign about half of
the total development costs of software to SQA of which \DDT s, i.e.,
analytical SQA, constitute the major part. Moreover, an understanding
of the economics is essential for project management to answer the
question how much quality assurance is enough. 
For example, Rai, Song, and Troutt \cite{rai98} state 
that a better understanding
of the costs and benefits should be useful to decision-makers.

However, the relationships regarding those costs and benefits are
often complicated and the data is difficult to obtain. Ntafos discusses
in \cite{ntafos01} that cost is a central factor but ``it is hard to
measure, data are not easy to obtain, and little has been done to
deal with it''. Nevertheless, there is a considerable amount of empirical
studies regarding \DDT s. The effectiveness and efficiency of testing
and inspections has been investigated intensively over the last decades.
Yet, we are not aware of a literature survey that summarises this
empirical knowledge with respect to an economics model.

\subsection{Problem}

The main practical problem is how we can optimally use \DDT s to improve
the quality of software. Hence, the two main issues are (1) in which order
and (2) with what effort the techniques should be used. This paper
concentrates on the subproblem that the collection of all relevant
data for a well-founded answer to these questions is not always possible.

\subsection{Contribution}

We review and summarise the empirical studies on various aspects of
\DDT s and software defects in general. The results of those studies
are assigned to the different input factors of an economics model of
analytical SQA. In particular, mean and median values of the input
factors are derived to allow an easier application of the model in
practice when not all factors are collectable. Furthermore, the
found distributions can be used in further analyses of the model.
Finally, the review reveals several areas that need further empirical
research.

\section{Software Quality Economics}
\label{sec:economics}

In this section, we introduce the general concept of quality costs for
software. Based on that, we describe an analytical, stochastic model
of the costs and benefits -- the economics -- of analytical SQA and
finally possibilities of its practical application.

\subsection{Software Quality Costs}
\label{sec:costs}

\emph{Quality costs} are the costs associated with preventing, finding,
and correcting defective work.
Based on experience from the manufacturing area
quality cost models have been developed explicitly for software.
These costs are divided into \emph{conformance} and \emph{nonconformance}
costs. 
The former comprises all costs that need to be spent to build the software in
a way that it conforms to its quality requirements. This can be further
broken down to \emph{prevention} and \emph{appraisal} costs. Prevention 
costs are for
example developer
training, tool costs, or quality audits, i.\,e.~costs for means to prevent
the injection of faults. The appraisal costs are caused by the usage of 
various types of tests and reviews.

\begin{figure}[h]
  \centering \includegraphics[width=.4\textwidth]{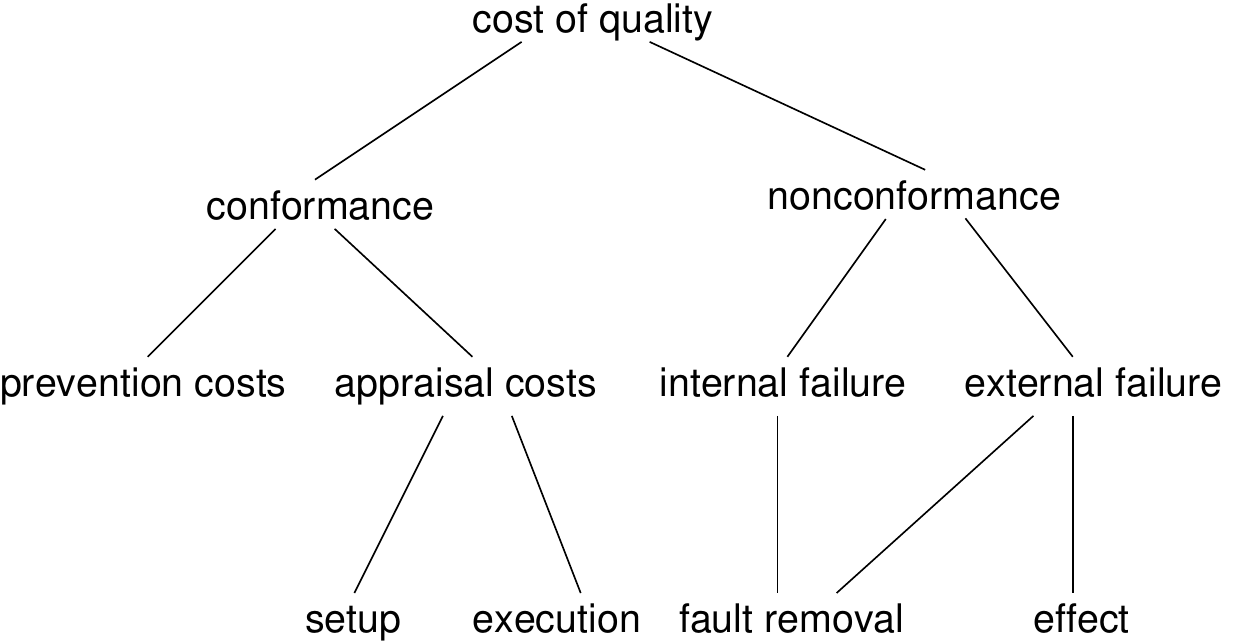}
  \caption{Overview over the costs related to quality}
  \label{fig:costs_overview}
\end{figure}

The \emph{nonconformance} costs come into play when the software does not
conform to the quality requirements. These costs are divided into
\emph{internal failure} costs and \emph{external failure} costs. The
former contains costs caused by failures that occur during development,
the latter describes costs that result from failures at the client.
A graphical overview is given in Fig.~\ref{fig:costs_overview}. Because
of the distinction between prevention, appraisal, and failure costs this
is often called \emph{PAF} model.
 
We add further detail to the PAF model
by introducing the main types of concrete costs that are important for
defect-detection techniques. Note that there are
more types that could be included, for example, maintenance costs.
However, we concentrate on a more reliability-oriented view.
The appraisal costs
are detailed to the \emph{setup} and \emph{execution} costs. The former
constituting all
initial costs for buying test tools, configuring the test environment,
and so on. The latter means all the costs that are connected to actual
test executions or review meetings, mainly personnel costs.

On the nonconformance side, we have \emph{fault removal} costs that can
be attributed to the internal failure costs as well as the external
failure costs. The reason is that the removal of a detected defect
always results in costs no matter whether it caused an internal
or external failure.

External failures also cause \emph{effect} costs. These are all further costs
with the failure apart from the removal costs. For example, compensation
costs could be part
of the effect costs, if the failure caused some kind of damage
at the customer site. We might also include further costs such as loss of 
sales because of bad
reputation in the effect costs but do not consider it explicitly
because it is out of scope of this paper.

\subsection{An Analytical Model}
\label{sec:model}

We give a short overview of an analytical model of \DDT s and refer
to \cite{wagner:issta06} for details. The model relates the discussed
cost factors and other technical factors with the aim to analyse the
economics of \DDT s. In particular, it can be used to plan the quality
assurance in a development project. Later we use the model as a basis
for reviewing the empirical literature and hence describe only briefly
the assumptions and equations.

\subsubsection{General}

We first describe an ideal model of quality economics
in the sense that we do not consider the practical use of the model but
want to mirror the actual relationships as faithfully as possible. 
We later simplify it for practical usages. The model is stochastic 
in the sense that it is based on expected values
as basis for decision making.

We divide the model in three main components:
\begin{itemize}
  \item Direct costs $\CDIRECT$
  \item Future costs $\CFUT$
  \item Revenues / saved costs $\REV$
\end{itemize}
The direct costs are characterised by containing only costs that
can be directly measured during the application of the technique. The
future costs and revenues are both concerned with the (potential) costs in
the field but can be distinguished because the future costs contain the
costs that are really incurred whereas the revenues are comprised of
saved costs.

We adapt the general notion of the difficulty of an application of 
technique $A$ to find
a specific fault $i$ from \cite{littlewood:tse00} denoted by $\theta_A (i)$
as a basic quantity for our model. In essence, it is the probability
that $A$ does not detect $i$. In the original definition this is
independent of time or effort but describes a ``single application''. 
We extend this using the length of the
technique application $t_A$. With length we do not mean calendar
time but effort measured in staff-days, for example, that was
spent for this technique application. Hence, the refined
difficulty function is defined as $\theta_A (i, t_A)$ denoting
the difficulty of $A$ detecting $i$ when applied with effort $t_A$.

Using this additional dimension we can also analyse different
functional forms of the difficulty functions depending
on the spent effort. This is similar to the informal curves shown by Boehm
\cite{boehm81} describing the effectiveness of different \DDT s
depending on the spent costs. Actually, the equations given for the model above
already contain that extended difficulty functions but they are not 
further elaborated. In \cite{wagner:issta06} we considered
several possible forms of the difficulty functions such as exponential
or linear.

We also assume that in the difficulty functions the concept of
defect classes is handled. A defect class is a group of defects
based on the document type it is contained in. Hence, we have
for each defect also its document class $c$, e.g., requirements
defects or code defects. This has especially an effect considering
that some techniques cannot be applied to all types of documents,
e.g., functional testing cannot reveal a defect in a design
document directly. It may however detect its successor in code.

This leads us to the further aspect that the defects occurring during
development are not independent. There are various dependencies
that could be considered but most importantly there is dependency
in terms of propagation.
Defects from earlier phases propagate to later phases and over
process steps. We actually do not consider the phases to be the important
factor here but the document types. In every development process there
are different types of documents, or artifacts, that are created.
Usually, those are requirements documents, design documents, code,
and test specifications. Then one defect in one of these documents
can lead to none, one, or more defects in later derived documents.

\subsubsection{Components}
\label{sec:equations_ideal}

We give an equation for each of the three components with respect
to single \DDT s first and later for a combination of techniques.
Note that the main basis of the model are expected values, i.e.,
we combine cost data with probabilities.

The direct costs are those costs that can be directly measured from
the application of a \DDT . They are dependent on the length $\LENGTH$
of the application.

From this we can derive the following definition for the
direct costs $\CDIRECT$:

\begin{equation}
  \label{eq:direct}
  \CDIRECT = \CSETUP + \CEXEC
             + \sum_{i}{
             (1 - \DIFF) \CREMV},
\end{equation}
where $\CSETUP$ are the setup costs, $\CEXEC$ the execution costs, and
$\CREMV$ the fault removal costs specific to that technique.

If some defects are not found, these will result in costs
in the future denoted by $\CFUT$. We divide these costs into 
the two parts fault removal
costs in the field $\CREMVF$ and failure effect costs $\CEFF$. The
latter contain all support and compensation costs as well as annoyed
customers as far as possible.

\begin{equation}
\label{eq:future}
  \CFUT = 
  \sum_i{\pi_i \DIFF (\CREMVF + \CEFF)},
\end{equation}
where $\pi_i = P$(fault $i$ is activated by
randomly selected
input and is detected and
fixed) \cite{littlewood:tse00}. Hence, it describes the probability that 
the defect leads to a failure
in the field.

It is necessary to consider not only costs with \DDT s but also revenues. 
These revenues
are essentially saved future costs. With each fault that we find in-house
we avoid higher costs in the future. Therefore, we have the same cost
categories but look at the faults that we find instead of the ones we
are not able to detect. We denote the revenues with $\REV$.
\begin{equation}
  \label{eq:saved}
  \REV =
  \sum_i{\pi_i (1 - \DIFF)(\CREMVF + \CEFF)}
\end{equation}
Because the revenues are saved future costs this equation looks
similar to Eq.~\ref{eq:future}. The difference is only that we
consider the faults that have not been found and hence use the 
probability of the negated difficulty, i.e., $1 - \DIFF$.

Typically, more than one technique is used to find defects.
The intuition behind that is that they find (partly) different defects.
These dependencies 
are often
ignored when the efficiency of \DDT s is analysed. Nevertheless, they
have a huge influence on the economics and efficiency. In our model
this is expressed using the different difficulty functions for specific
faults and techniques.

For the direct costs it means that we sum over all different
applications of \DDT s. We define that
$X$ is the ordered set of the applied \DDT s. In each application we
use Eq.~\ref{eq:direct} with the extension that we not only take the
probability that the technique finds the fault into account but 
also that the ones
before have not detected it. Here also the defect propagation
needs to be considered, i.e., that not only the defect itself has
not been detected but also its predecessors $R_i$.

For the sake of readability we introduce the abbreviation $\Theta(x, R_i)$
for the
probability that a fault and its predecessors have not been found
by previous -- before $x$ -- applications of \DDT s.
\begin{equation}
\Theta(x, R_i) = \prod_{y < x}{\biggl[\theta_y(i,t_y)}
                         \prod_{j \in R_i}{\theta_y(j,t_y)\biggr]},
\end{equation}
hence, for each technique $y$ that is applied before $x$ we multiply
the difficulty for the fault $i$ and all its predecessors as
described in the set $R_i$. The combined direct costs $d_X$ of
a sequence of \DDT\ applications $X$ is then defined as follows:
\begin{equation}
\label{eq:direct_total}
  d_X = \sum_{x \in X}{\biggl[ u_x + e_x(t_x)} + 
        \sum_i{\Bigl( (1 - 
         \theta_x(i,t_x))}
         \Theta(x, R_i) \Bigr) 
         \CREMVX \biggr]
\end{equation}

The equation for the revenues $r_X$ of several technique applications
$X$ uses again a sum over all technique
applications. In this case we look at the faults that occur, that are
detected by a technique and neither itself nor its predecessors
have been detected by the earlier
applied techniques.

\begin{equation}
\label{eq:revenues_total}
  r_X = \sum_{x \in X}{\sum_i{\biggl[ \Bigl( \pi_i (1 - \theta_x(i,t_x))
         \Theta(x, R_i)}}
         \Bigr)
         \bigl(
         \CREMVF + \CEFF \bigr)  \biggr]
\end{equation}

The total future costs are simply the costs of each fault with the
probability that it occurs and all techniques failed in detecting it
and its predecessors. In this case, the abbreviation $\Theta(x, R_i)$
for accounting of the effects of previous technique applications
cannot be directly used because the outermost sum is over all the
faults and hence the probability that a previous technique detected
the fault is not relevant. The abbreviation $\Theta'(x, R_i)$
that describes only the product of the difficulties of detecting
the predecessors of $i$ is hinted in the following equation for
the future cost $o_X$ of several technique applications $X$.

\begin{equation}
\label{eq:future_total}
  o_X = \sum_i{\biggl[ \pi_i \prod_{x \in X}{\theta_x(i,t_x)}}
        \underbrace{\prod_{y < x}{\prod_{j \in R_i}
                   {\theta_y(j,t_y)}}}_{\Theta'(x, R_i)}
        (\CREMVF + \CEFF) \biggr]
\end{equation}

\subsubsection{ROI}

One interesting metric based on these values is the \emph{return on investment}
(ROI) of the \DDT s. The ROI -- also called \emph{rate of return} -- 
is commonly defined as the gain divided
by the used capital. Boehm et al.\ \cite{boehm04} use the equation
$(\mbox{Benefits} - \mbox{Costs})/\mbox{Costs}$. To calculate 
the total ROI with our model we have to use 
Eqns.~\ref{eq:direct_total}, \ref{eq:future_total}, and 
\ref{eq:revenues_total}.

\begin{equation}
  \mbox{ROI} = \frac{r_X - d_X - o_X}{d_X + o_X}
\end{equation}

This metric can be used for two purposes: (1) an up-front evaluation
of the quality assurance plan as the \emph{expected} ROI of performing
it and (2) a single post-evaluation of the quality
assurance of a project. In the second case we can substitute the
initial estimates with actually measured values. However, not all of the factors
can be directly measured. Hence, also the post evaluation metric
can be seen as an \emph{estimated} ROI.

\subsection{Practical Model}
\label{sec:practical}

The ideal model can be used for theoretical analyses
but is too detailed for a practical application. Hence, a simplified
version of this model is available that can be used to plan the
quality assurance of a development project using historical project
data. Details can be found in \cite{wagner:issta06}. We only
describe the additional assumptions and simplifications in the
following.

For the simplification of the model, we use the following 
additional assumptions:

\begin{itemize}
  \item Faults can be categorised in useful defect types.
  \item Defect types have specific distributions regarding
        their detection difficulty, removal costs, and failure probability.
  \item The linear functional form of the difficulty approximates
        all other functional forms sufficiently.
\end{itemize}

We define $\TYPE$ to be the defect type of fault $i$. It is determined
using the defect type distribution of older projects.
In this way we do not have to look at individual
faults but analyse and measure defect types for which the determination
of the quantities is significantly easier. 

In the practical model we assumed that the defects can be grouped in
``useful'' classes or defect types. For reformulating the equation
it was sufficient to consider the affiliation of a defect to a type
but for using the model in practice we need to further elaborate on
the nature of defect types and how to measure them.

We also lose the concept of defect propagation as it was shown not to
have a high priority in the analyses above but it introduces significant
complexity to the model. Hence, the practical model can be simplified
notably.

\section{Empirical Knowledge}
\label{sec:empirical}

We review and summarise the empirical knowledge available for the
quality economics of defect-detection techniques introducing the
approach in general and then describing the relevant studies
and results for each
of the model factors for different types of techniques and
defects in general.
We can only give summaries here and refer to \cite{wagner:tumi06}
for details.

\subsection{General}

The field of quality assurance and \DDT s in particular
has been subject to a number of empirical studies over
the last decades. These studies were used to assess specific techniques
or to validate certain laws and theories about defect-detection.
Our focus lies on the economic relationships in the following.

\subsubsection{Approach}

This survey aims at reviewing and summarising the existing empirical
work that can be used to approximate the input parameters of the
economics model proposed in Sec.~\ref{sec:model}. For this we take
all officially published sources into account, i.e.\ books, journal
articles, and papers in workshop and conference proceedings. In
total we reviewed 68 papers mainly following references
from existing surveys and complementing those with newer publications.
However, note that we only included studies with data relevant for
the economics model. In particular, studies only with a comparison
of techniques without detailed data for each were not taken into account.

In the literature review in the following sections, we structure the 
available work in three parts
for dynamic testing, review and inspection, and static analysis tools.
We give a short characterisation for each category and describe
briefly the available results for each relevant model input factor.
We prefer to use and cite detailed results of single applications
of techniques but also take mean values into account if necessary.
We also summarise the combination of the results in terms of the
lowest, highest, mean, and median value for each input factor.

We deliberately refrain from assigning weights to the various
values we combine although some of them are from single experiments
while others represent average values. The reason is that we often
lack knowledge on the sample size used and either we would estimate
it or ignore the whole study result. An estimate of the sample size
would introduce additional blurring into the data and omitting data
considering the limited amount of data available is not advisable.
Hence, we assume each data set of having equal weight.

\subsubsection{Difficulty}
\label{sec:difficulty}

The difficulty function $\theta$ is hard to determine because it
is complex to analyse the difficulty of finding each potential fault
with different \DDT s. Hence, we need to use the available empirical
studies to get reasonable estimates. Firstly, we can use the numerous
results for the effectiveness of different test techniques. The
effectiveness is the ratio of found defects to total defects and hence
in some sense the counterpart to the difficulty function. In the
paper of Littlewood et al.\ \cite{littlewood:tse00}, where the idea
of the difficulty function originated, \emph{effectiveness} is actually
defined as
\begin{equation}
1 - E_{p^*}(\theta_A(i)),
\end{equation}
where $E_{p^*}$ denotes a mean obtained with respect to the probability
distribution $p^*$, i.e.\ the probability distribution of the presence
of faults. 

As a first, simple approximation we then define the following for
the difficulty functions.

\begin{equation}
\theta_A = 1 - \mbox{effectiveness}
\end{equation}

The problem is that this is really a coarse-grained approximation
that does not reflect the diversity of defect detection of different
techniques. Hence, we also need to analyse studies that use different
defect types in the sense of the practical model from 
Sec.~\ref{sec:practical}.

\subsection{Dynamic Testing}
\label{sec:tests}

The first category of \DDT s we look at is also the most important
one in terms of practical usage. Dynamic testing is a technique that
executes software with the aim to find failures.

\subsubsection{Classification}
\label{sec:tests_class}

\begin{figure}[h]
\begin{center}
  \includegraphics[width=.4\textwidth]{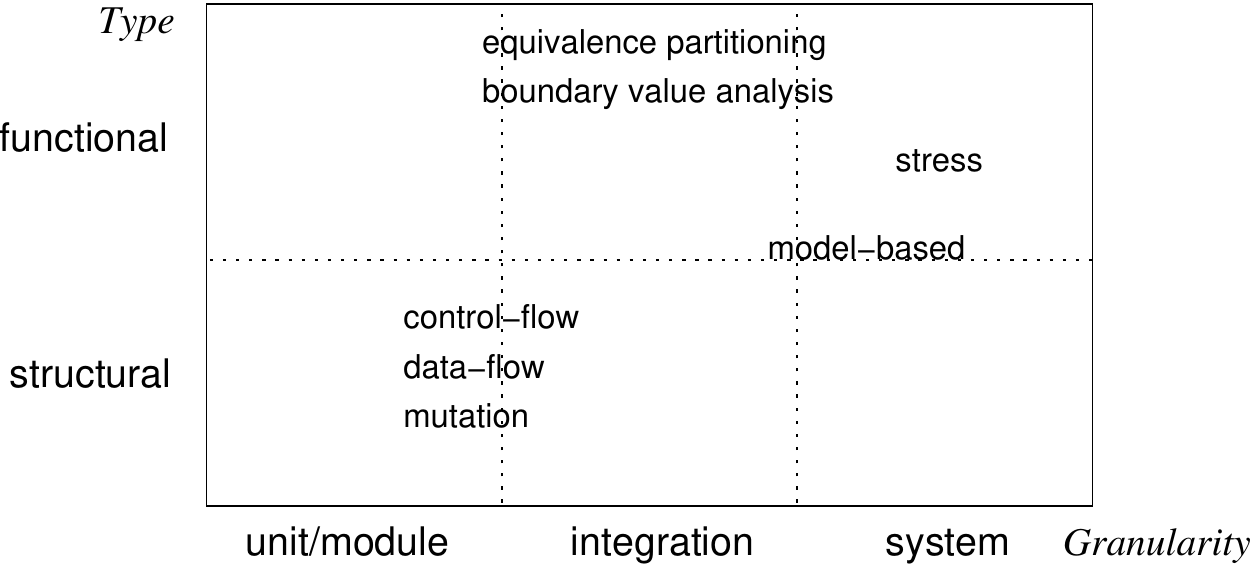}
  \caption{The two basic dimensions of test techniques}
  \label{fig:test_dimensions}
\end{center}
\end{figure}

There are various possibilities to classify different test techniques.
One can identify at least two dimensions to structure the techniques.
(1) The granularity of the test object and (2) the test case derivation
technique. Fig.~\ref{fig:test_dimensions} shows these two dimensions
and contains some concrete examples and how they can be placed according
to these dimensions.

The types of test case derivation can be divided on the top level into
(1) functional and (2) structural test techniques. The first only uses
the specification to design tests, whereas the latter relies on the
source code and the specification. In functional testing generally
techniques such as equivalence partitioning and boundary value analysis
are used. Structural testing is often divided into control-flow and
data-flow techniques. For the control-flow coverage metrics such as
statement coverage or condition coverage are in use. The data-flow
metrics measure the number and types of uses of variables. 

On the granularity dimension we normally see the phases unit, module
or component test, integration test, and system test. In unit
tests only basic components of the system are tested using stubs to
simulate the environment. During integration tests the components
are combined and their interaction is analysed. Finally, in system
testing the whole system is tested, often with some similarity
to the later operational profile. This also corresponds to the
development phases. Hence, the granularity dimension can also be
seen as phase dimension.

\subsubsection{Setup and Execution Costs}

We look at the setup and execution costs in more detail in the
following. For both cost types the empirical data is limited.
However, this is not a great problem because this data
can be easily collected in a software company during projects.

The setup costs are mainly the staff-hours needed for understanding
the specification in general and setting up
the test environment. For this we can use data from 
\cite{jones91}. There the typical setup effort is given in relation
to the size of the software measured in function points (fp).
Unit tests need 0.50 h/fp, function tests
0.75 h/fp, system test 1.00 h/fp, and field tests 0.50 h/fp.
We have no data for typical costs of tools and hardware but this
can usually be found in accounting departments when using the
economics model in practice.

In the case of execution costs its
even easier than for setup costs as apart from the personnel costs
all other costs can be neglected. One could include costs such as
energy consumption but they are extremely small compared to the
costs for the testers. Hence, we can reduce this to the typical,
average costs for the staff. However, we also have average values
per function point from
\cite{jones91}. There the average effort for unit tests is 0.25 h/fp, 
for function tests,
system tests, and field tests 0.50 h/fp.

\subsubsection{Difficulty}

As discussed in Sec.~\ref{sec:difficulty}, there are nearly no
studies that present direct results for the difficulty function
of \DDT s. Hence, we analyse the effectiveness and efficiency
results first. Those are dependent on the test case derivation
technique used.

In the following we summarise a series of studies that have been
published regarding the effectiveness of testing in general and
specific testing techniques.

A summary of the found effectiveness of functional and structural
test techniques can be found in Tab.~\ref{tab:tests_sum_effectiveness}.
We can observe that the mean and median values are all quite close which
suggests that there are no strong outliers. However, the range in general
is rather large, especially when considering all test techniques.
When comparing functional and structural testing, there is no significant
difference visible.

\begin{table}[h]
\caption{Summary of the effectiveness of test techniques (in percentages)}
\label{tab:tests_sum_effectiveness}
\begin{center}
\begin{tabular}{|l|rrrr|}
\hline
Type & Lowest & Mean & Median & Highest\\
\hline
Functional & 33 & 53.26 & 48.85 & 88\\
Structural & 17 & 54.78 & 56.85 & 89\\
All        & 7.2 & 49.85 & 47 & 89\\
\hline
\end{tabular}
\end{center}
\end{table}

The effectiveness gives a good first approximation of $\theta$. The
efficiency measures the number of detected defects per effort unit
(staff-hours, for example). It cannot be used directly in the economics
model but is also summarised as it is an important metric itself.

The found efficiencies of functional and structural
test techniques are summarised in Tab.~\ref{tab:tests_sum_efficiency}.
We assume that one staff-hour consists of 60 staff-minutes. The results
show that the data is quite homogenous because the means and medians
are all equal or nearly equal and the ranges are rather small.
Especially for functional testing it is only slightly above 1 defect/hour
difference between the lowest and the highest value.

\begin{table}[h]
\caption{Summary of the efficiency of test techniques (in defects
         per staff-hour)}
\label{tab:tests_sum_efficiency}
\begin{center}
\begin{tabular}{|l|rrrr|}
\hline
Type & Lowest & Mean & Median & Highest\\
\hline
Functional & 1.22 & 1.72 & 1.71 & 2.47\\
Structural & 0.22 & 1.5 & 2.07 & 2.2\\
All        & 0.04 & 1.26  & 1.5 & 2.47\\
\hline
\end{tabular}
\end{center}
\end{table}

The first approximation of the difficulty
functions is given in Tab.~\ref{tab:tests_diff_first}. We used the
results of the effectiveness summary above. Hence, the observations
are accordingly.

\begin{table}[h]
\caption{First approximation of the difficulty functions for testing}
\label{tab:tests_diff_first}
\begin{center}
\begin{tabular}{|l|rrrr|}
\hline
Type & Lowest & Mean & Median & Highest\\
\hline
Functional & 12 & 46.74 & 51.15 & 67\\
Structural & 11 & 45.22 & 43.15 & 83\\
All        & 11 & 50.15 & 53 & 92.8\\
\hline
\end{tabular}
\end{center}
\end{table}

Finally, for a real application to our economics model we need
to differentiate between different fault types. Basili and Selby
analysed the effectiveness of functional and structural testing
regarding different defect types in \cite{basili87}.
Tab.~\ref{tab:eff_types} shows the derived difficulties using
the first approximation.

\begin{table}[h]
\caption{Difficulties of functional and structural testing
         for detecting different defect types}
\label{tab:eff_types}
\begin{center}
\begin{tabular}{|l|rrr|}
\hline
       & Functional & Structural & Overall\\
       & Testing & Testing & \\
\hline
Initial.  & 25.0 & 53.8 & 38.5\\
Control   & 33.3 & 51.2 & 47.2\\
Data      & 71.7 & 73.2 & 74.7\\
Computat. & 35.8 & 41.2 & 75.4\\
Interface & 69.3 & 75.4 & 66.9\\
Cosmetic  & 91.7 & 92.3 & 89.2\\
\hline
\end{tabular}
\end{center}
\end{table}

It is obvious that there are differences of the two techniques
for some defect types, in particular initialisation and control
defects. As we are only aware of this single study it is difficult
to generalise the results.

\subsubsection{Removal Costs}

The removal costs are dependent on the second dimension of
testing (cf.\ Sec.~\ref{sec:tests_class}): the phase in which
it is used. This is in general a very common observation in
defect removal that it is significantly more expensive to fix
defects in later phases than in earlier ones. Specific for testing,
in comparison with static techniques, is that defect removal not only
involves the act of changing the code but before that of localising
the fault in the code. This is simply a result of the fact that testing
always observes failures for which the causing fault is not
necessarily obvious. We cite the
results of several studies regarding those costs in the following.

Some statistics of the data above on the removal costs are summarised
in Tab.~\ref{tab:tests_sum_removal}. We assume a staff-day to consist
of 6 staff-hours and combined the functional and system test phases
into the one phase ``system test''. The removal costs (or efforts) of
the three phases can be given with reasonable results. A combination
of all values for a general averages does not make sense as we get a
huge range and a large difference between mean and median. This suggests
a real difference in the removal costs over the different phases which
is expected from standard software engineering literature.

\begin{table}[h]
\caption{Summary of the removal costs of test techniques (in staff-hours
         per defect)}
\label{tab:tests_sum_removal}
\begin{center}
\begin{tabular}{|l|rrrr|}
\hline
Type & Lowest & Mean & Median & Highest\\
\hline
Unit        & 1.5 & 3.46 & 2.5 & 6\\
Integration & 3.06 & 5.42 & 4.55 & 9.5\\
System      & 2.82 & 8.37 & 6.2 & 20\\
All         & 0.2  & 8 & 4.95 & 52\\
\hline
\end{tabular}
\end{center}
\end{table}

\subsection{Review and Inspection}
\label{sec:reviews}

The second category of \DDT s under consideration are reviews
and inspections, i.e.\ document reading with the aim to improve them.

\subsubsection{Classification}

We use the term \emph{inspection} here in a broad sense for all
kinds of document reading with the aim of defect-detection. In
most cases \emph{review} is used interchangeably. We can
then identify differences mainly in the technical dimension, e.g., 
in the process of the inspections, for example
whether explicit preparation is required. Other differences lie in the
used reading techniques, e.g.\ checklists, in the required roles, or
in the products that are inspected.

A prominent example is the formal or Fagan inspection that has a
well-defined process with a separate preparation and meeting and defined
roles. Another often used technique is the walkthrough. In this technique
the moderator guides through the code but no preparation is required.

\subsubsection{Setup and Execution Costs}

The first question is whether reviews and inspections
do have setup costs. We considered those costs to be fixed and independent
of the time that the \DDT\ is applied. In inspections we typically have
a preparation and a meeting phase but both can be varied in length to
detect more defects. Hence, they cannot be part of the setup costs.
However, we have also an effort for the planning and the kick-off that
is rather fixed. We consider those as the setup costs of inspections. 
One could also include costs for printing the documents but these costs
can be neglected.
Grady and van Slack describe in \cite{grady94} the experience of
Hewlett-Packard with inspections. They give typical time effort for
the different inspection phases, for planning 2 staff-hours and for
the kick-off 0.5 staff-hours.

The execution costs are for inspections
and reviews only the personnel costs as long as there is no supporting
software used. Hence, the execution costs are dependent on the factor
$t$ in our model. Nevertheless, there are some typical values for
the execution costs of inspections.

We can derive some LOC-based statistics
We assume for the sake of simplicity
that all used varieties of the LOC metric are approximately equal. The
results are summarised in Tab.~\ref{tab:reviews_sum_execution}. The mean
and median values are all close. Only in code inspection meetings, there
is a difference which can be explained by the small sample size. Note also
that there is a significant difference between code and design inspections
as the latter needs on average only half the execution costs. This might
be explained by the fact that design documents are generally more abstract
than code and hence easier to comprehend.

\begin{table}[h]
\caption{Summary of the execution costs of inspection techniques (in staff-hours
         per KLOC)}
\label{tab:reviews_sum_execution}
\begin{center}
\begin{tabular}{|l|rrrr|}
\hline
Design & Lowest & Mean & Median & Highest\\
\hline
Preparation & 3.6 & 4.68 & 4.68 & 5.76\\
Meeting     & 3.6 & 4.07 & 4.07  & 4.54\\
All         & 7.2 & 8.75 & 8.75  & 10.3\\
\hline
Code & Lowest & Mean & Median & Highest\\
\hline
Preparation & 4.91 & 6.49 & 6.67 & 7.9\\
Meeting     & 3.32 & 7.02 & 4.4  & 13.33\\
All         & 6.67  & 13.2 & 11.15  & 22\\
\hline
\end{tabular}
\end{center}
\end{table}

Moreover, note 
that many authors
give guidelines for the optimal inspection rate, i.e.\ how fast the inspectors
read the documents. This seems to have an significant impact on the
efficiency of the inspection. For example, in \cite{gilb93} the optimal bandwidth 
of the inspection rate is $1 \pm 0.8$ pages per hour where one page contains 300 words.
As other authors give similar figures,
we can summarise this easily with saying that the optimal inspection
rate lies about one page per hour. However, the effect of deviation from
this optimum is not well understood. This, however, would increase the
precision of models such as the one presented in Sec.~\ref{sec:model}.

\subsubsection{Difficulty}

Similar to the test techniques we start with
analysing the effectiveness of inspections and reviews that is later used
in the approximation of the difficulty.

We also summarise these results using the lowest, highest, mean, and 
median value in Tab.~\ref{tab:reviews_sum_effectiveness}. We observe
a quite stable mean value that is close to the median with about 30\%.
However, the range of values is huge. This suggests that an inspection
is dependent on other factors to be effective.

\begin{table}[h]
\caption{Summary of the effectiveness of inspection techniques (in 
         percentage)}
\label{tab:reviews_sum_effectiveness}
\begin{center}
\begin{tabular}{|rrrr|}
\hline
Lowest & Mean & Median & Highest\\
\hline
8.5 & 34.14 & 30 & 92.7\\
\hline
\end{tabular}
\end{center}
\end{table}

The efficiency relates the effectiveness with the
spent effort. Again, this is not directly usable in the analytical
model but nevertheless can give further insights into the relationships
of factors.

The statistics for the efficiency of reviews and inspections can
be found in Tab.~\ref{tab:reviews_sum_efficiency}. We do not 
distinguish different processes and reading techniques here because
then we would not have enough information on these in most studies.
The mean and median are close, therefore the data set is reasonable.
We also observe a large range from 0.16 to 6 defects/staff-hour.

\begin{table}[h]
\caption{Summary of the efficiency of inspection techniques (in 
         defects per staff-hour)}
\label{tab:reviews_sum_efficiency}
\begin{center}
\begin{tabular}{|rrrr|}
\hline
Lowest & Mean & Median & Highest\\
\hline
0.16 & 1.87 & 1.18 & 6\\
\hline
\end{tabular}
\end{center}
\end{table}

Using the first, simple approximation, we can
derive statistics for the difficulty of inspections in reviews in
Tab.~\ref{tab:reviews_sum_difficulty}.

\begin{table}[h]
\caption{Average difficulty of inspections (in percentages)} 
\label{tab:reviews_sum_difficulty}
\begin{center}
\begin{tabular}{|rrrr|}
\hline
Lowest & Mean & Median & Highest\\
\hline
7.3 & 65.86 & 70 & 91.5\\
\hline
\end{tabular}
\end{center}
\end{table}

Analogous to the test techniques, we only have
one study about effectiveness and defect types \cite{basili87}.
The derived difficulty functions are given in 
Tab.~\ref{tab:reviews_diff_types}. Also for inspections large
differences between the defect types are visible but a single study
does not guarantee generalisability.

\begin{table}[h]
\caption{Difficulty of inspections to find different defect types
\label{tab:reviews_diff_types}}
\begin{center}
\begin{tabular}{|l|r|}
\hline
Initial.  & 35.4 \\
Control   & 57.2 \\
Data      & 79.3 \\
Computat. & 29.1 \\
Interface & 53.3 \\
Cosmetic  & 83.3 \\
\hline
\end{tabular}
\end{center}
\end{table}

\subsubsection{Removal Costs}

The summary of the the removal costs can be found in 
Tab.~\ref{tab:reviews_sum_removal}. For the design reviews a
strong difference between the mean and median can be observed.
However, in this case this is not because of outliers in the
data but because of the small sample size of only four data points.

\begin{table}[h]
\caption{Summary of the removal costs of inspections (in staff-hours
         per defect)}
\label{tab:reviews_sum_removal}
\begin{center}
\begin{tabular}{|l|rrrr|}
\hline
Phase & Lowest & Mean & Median & Highest\\
\hline
Requirements & 0.05 & 1.06 & 1.1 & 2\\
Design       & 0.07 & 2.31 & 0.83 & 6.3\\
Coding       & 0.17 & 2.71 & 1.95 & 6.3\\
All          & 0.05 & 1.91 & 1.2  & 7.5\\
\hline
\end{tabular}
\end{center}
\end{table}

\subsection{Static Analysis Tools}
\label{sec:static}

The third and final category is tool-based analysis of software
code to automatise the detection of certain types of defects.

\subsubsection{Classification}

The term \emph{static analysis tools} denotes a huge field of
software tools that are able to find (potential) defects in
software code without executing it. 
Those analysis tools use various techniques to
identify critical code pieces. The most common one is to define typical
bug patterns that are derived from experience and published common pitfalls
in  a certain programming language. Furthermore, coding guidelines and standards
can be checked to allow a better readability. Also, more sophisticated
analysis techniques based on the dataflow and controlflow are used. Finally,
additional annotations in the code
are introduced by some tools \cite{flanagan02} to allow an extended static
checking and a combination with model checking.

The results
of such a tool
are, however, not always real defects but can be seen as a warning that a
piece of code is critical in some way. Hence, the analysis with respect
to true and false positives is essential in the usage of bug finding tools.

There are only few studies about static analysis tools and
hence we can only present limited empirical knowledge.

\subsubsection{Setup and Execution Costs}

There are no studies with data about the setup and execution costs
of using static analysis tools. Still, we try to analyse those
costs and their influence in the context of such tools. 

The setup
costs are typically quite small consisting only of (possible) tool
costs --- although there are several freely available tools --- and
effort for the installation of the tools to have it ready for analysis.

The execution costs are small in the first step because we only need
to select the source files to be checked and run the automatic analysis.
For tools that rely on additional annotations the execution costs
are considerably higher. The second step, to distinguish between true
and false positives, is much more labour intensive than the first step.
This requires possibly to read the code and analyse the interrelationships
in the code which essentially constitutes a reviews of the code. Hence,
the ratio of false positives is an important measure for the
efficiency and execution costs of a tool.

In \cite{wagner:testcom05} we found that the average ratio of false
positives over three tools for Java was 66\% ranging from 31\% up
to 96\%. In \cite{johnson04} a static analysis tools for C code 
is discussed. The
authors state that sophisticated analysis of, for example, pointers leads
to far less false positives than simple syntactical checks.

\subsubsection{Difficulty}

Static analysis techniques are evaluated in \cite{howden78}.
Interface consistency rules and anomaly analysis revealed 2 and 4
faults of 28, respectively.
We also analysed the effectiveness of three Java bug finding tools in
\cite{wagner:testcom05}. After eliminating the false positives, the
tools were able to find 81\% of the known defects over several projects.
However, the defects had mainly a low severity. For the severest defects
the effectiveness reduced to 22\%, for the second severest defects even
to 20\%. For lower severities the effectiveness lies between 70\% -- 88\%.

\subsection{Defects}

In this section we look at the quantities that are independent from
a specific \DDT\ and can be associated to defects. We are interested
in typical defect type distributions, removal costs in the field,
failure severities for the calculation of possible effect costs,
and failure probabilities of faults.

\subsubsection{Defect Introduction}

The general probability that a specific
possible fault is introduced into a specific program cannot be determined
in general without replicated experiments. However, we can give some
information when considering defect types. We can determine the defect
type distribution for certain application types. Yet, there is only little
data published. Sullivan and Chillarege described the defect type
distribution of the database systems DB2 and IMS in \cite{sullivan92}.
Most of the defects were in assignment checking, data structures, and
algorithm. Interface and timing defects constitute only a small share
of the total number of defects.

Lutz and Mikulski used for defects in NASA software a slightly different
classification of defects in \cite{lutz04} but they also have
algorithms and assignments as types with a lot of occurrences. The most
often defect type, however, is \emph{procedures} meaning missing procedures
or wrong call of procedures.

In \cite{rubey75} types and severities of software defects
are described. We can observe that logical and data
access defects account for most of the serious defects.
Furthermore, most of the defects were defects in the specification.

As a summary, we can formulate that the defect types are strongly
domain- and problem-specific and general conclusions are hard to make.

\subsubsection{Removal Costs} 

In this section we analyse only the removal
costs of defects in the field as during development we consider the
removal costs to be dependent on the used \DDT .

\begin{figure}[h]
  \centering \includegraphics[width=.4\textwidth]{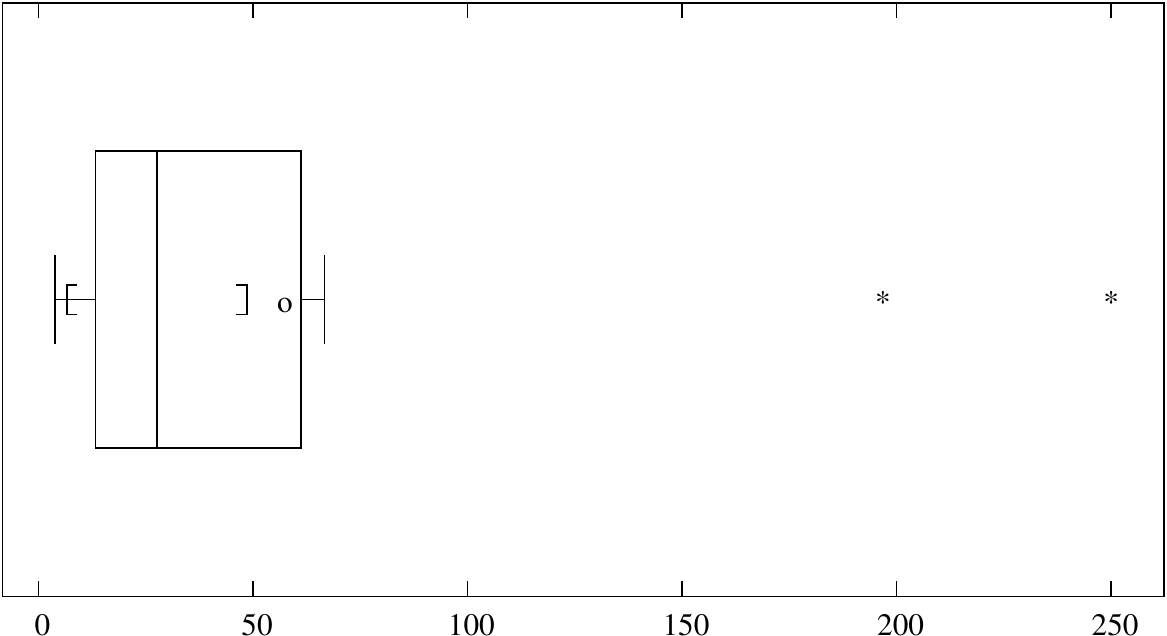}
  \caption{Box plot of the removal costs of field defects in staff-hours
           per defect}
  \label{fig:defect_removal_box}
\end{figure}

For the removal costs we have enough data to give reasonably some
statistics in Tab.~\ref{tab:defect_sum_removal}. Note that in this
summary the mean and median are extremely different. The mean is more
than twice the median. This indicates that there are outliers in the
data set that distort the mean value. Hence, we look at a box plot
of the data in Fig.~\ref{fig:defect_removal_box}.

\begin{table}[h]
\caption{Summary of the removal costs of field defects (in staff-hours
         per defect)}
\label{tab:defect_sum_removal}
\begin{center}
\begin{tabular}{|rrrr|}
\hline
Lowest & Mean & Median & Highest\\
\hline
3.9 & 57.42 & 27.6 & 250\\
\hline
\end{tabular}
\end{center}
\end{table}

The box plot in Fig.~\ref{fig:defect_removal_box} shows two strong
outliers that we can eliminate to get a more reasonable mean value.
With the reduced data set we get a mean value of 27.24 staff-hours per
defect and a median of 27 staff-hours per defect. Hence, we have a more
balanced data set with a mean value that can be further used.

\subsubsection{Effect Costs}

The effect costs are probably the most difficult ones to obtain.
One reason is that these are highly domain-specific. Another is
that companies often do not publish such data as it could influence
their reputation negatively. There is also one more inherent problem.
It is often just not possible to to assign such costs to a single
software fault. The highly complex configurations and the combination
with hardware and possibly mechanics of software make such an assignment
extremely difficult. 

Yet, we cite two studies that published distribution of severity
levels of defects. We consider the severity as the best available
substitute of effect costs because more severe defects are probably
more costly in that sense. However, this leaves us still with the
need of a mapping of severity levels with typical effect costs.

Jones \cite{jones91} states that the
        typical severity levels (1: System or program inoperable, 2: Major
        functions disabled or incorrect, 3: Minor functions disabled or 
        incorrect,
        4: Superficial error) have the following distribution:
        \begin{enumerate}
        \item 10\% or 3\%
        \item 40\% or 15\%
        \item 30\% or 60\%
        \item 20\% or 22\%

        \end{enumerate}
In \cite{christmansson96} is reported that six error types accounted for
        nearly 80\% of the highest severity defects. Nine error types accounted
        for about 80\% of the defects exposed by recovery procedures or
        exception handlers. They used ODC for classification.

\subsubsection{Failure Probability}

The failure probability of a fault is also one of the most difficult
parts to determine in the economics model. Although there is the whole
research field of software reliability engineering, there are only few
studies that show representative distribution of such probabilities.
The often cited paper from Adams
\cite{adams84} is one of the few exceptions. He mainly shows that
the failure probabilities of the faults have an underlying geometric
progression. This observation was also made in NASA studies reported
in \cite{nagel84}.

This relationship can also be supported by data from Siemens when
used in a reliability model \cite{wagner:ada-europe06}. The geometric
progression fitted reasonably on all analysed projects.

\section{Discussion}
\label{sec:discussion}

Some of the summaries allow a comparison over different
techniques. Most interestingly, the difficulty of finding
defects is different between tests and inspections with
inspections having more difficulties. Tests tend on average
to a difficulty of 0.45 whereas inspections have about 0.65.
The static analysis tools are hard to compare because of the
limited data but seem to be better in total but much worse
considering severe defects.

The removal costs form a perfect series over the various
techniques. As expected, the requirements reviews only need
about 1 staff-hour of removal effort which rises over the
other reviews to the unit tests with about 3.5 staff-hours.
Over the testing phases we have again an increase to the
system test with about 8 staff-hours. The field defects
are then more than three times as expensive with 27 staff-hours.
Hence, we can support the typical assumption that it gets
more and more expensive to remove a defect over the development
life-cycle.

We are aware that this survey can be criticised in many ways. One
problem is clearly the combination of data from various sources without
taking into account all the additional information. However, the aim
of this survey is not to analyse specific techniques in detail and
statistically test hypotheses but to determine some average values, some
rules of thumb as approximations for the usage in an economics model.
Furthermore, for many studies we do not have enough information for
more sophisticated analyses.

Jones gives in \cite{jones91}
a rule of thumb: companies 
that have testing departments staffed by
trained specialists will average about 10 to 15 percent higher in
cumulative testing efficiency than companies which attempt testing by
using their ordinary programming staff. Normal unit testing by programmers
is seldom more than 25 percent efficient and most other forms of testing
are usually less than 30 percent efficient when carried out by untrained
generalists. A series of well-planned tests by a professionally staffed
testing group can exceed 35 percent per stage, and 80 percent in overall
cumulative testing efficiency.
Hence, the staff experience can be seen as one of the influential
factors on the variations in our results.

\section{Related Work}
\label{sec:related}

The complete analytical model as described in this paper was
published in \cite{wagner:issta06}.

The available related work to the economics model
can generally classified in two categories:
(1) theoretical models of the effectiveness and efficiency of either
test techniques or inspections and (2) economic-oriented, abstract
models for quality assurance in general. The first type of models is
able to incorporate interesting technical details but are typically
restricted to a specific type of techniques and often economical
considerations are not taken into account. The second type of models
typically comes from more management-oriented researchers that
consider economic constraints and are able to analyse different types
of defect-detection but often deal with the technical details in a
very abstract way. Because of the limited space we do not cite those
studies but refer to \cite{wagner:issta06} for details.

There are also already some literature surveys on \DDT s.
Juristo et al.\ summarise in \cite{juristo04} the main experiments
regarding testing techniques of the last 25 years. Their main focus
is to classify the techniques and experiments and compare the techniques
but not to collect and compare actual figures.
In \cite{briand98} several sources from the
literature for inspection efficiency were used to build efficiency
benchmarks. Laitenberger published a survey on inspection technologies
in \cite{laitenberger02a}. He also included data on effectiveness and
effort but without relating it to a model.

\section{Conclusions}
\label{sec:conclusion}

We summarise the main results and contribution of the paper in the
following and give directions for further research.

\subsection{Summary}

We reviewed and summarised the relevant empirical studies on
\DDT s that can be used to determine the input factors of an
economics model of software quality assurance. The results of the studies
were structured with respect to the technique they pertained and the
corresponding input factor of the model. The difficulty function is
the most complex factor to determine. We introduced two 
methods to obtain approximation of the factors for the three groups
of techniques.

We observed that test techniques tend to be more efficient in
defect detection having lesser difficulties but to have larger
removal costs. A further analysis in the model might reveal
which factor is more important. Furthermore, the removal costs
increase also strongly considering different types of tests
or reviews, i.e., during unit tests fault removal is considerably
cheaper than during system tests. This suggests that unit-testing
is very cost-efficient.

\subsection{Further Research}

We discussed an optimal inspection rate, i.e., the optimal effort per
LOC regarding the efficiency of the inspection, and noted that it is
not well understood how a deviation from this optimal rate has effects
on other factors in defect detection. Hence, further studies and experiments
on this would be needed to refine the economics model and improve the
analysis and prediction of the optimal quality assurance.

The difficulty of detecting different defect types with different
detection techniques should be investigated more thoroughly. The 
empirical knowledge
is extremely limited there although this would allow an improved
combination of diverse techniques.

The effect costs are a difficult part of the failure costs. They are
a highly delicate issue for most companies. Nevertheless, empirical
knowledge is also important there to be able to estimate the influence
on the total quality costs.

The collected empirical knowledge on the input factors 
can be used to refine the sensitivity analysis of the model that was
done in \cite{wagner:issta06}. A sensitivity analysis can be used to
identify the most important input factors and their contribution to
the variation in the output. The mean value and knowledge
on the distribution (if available) can be used to generate more 
accurate input data
to the analysis.

We will also extend the economics models by a size metric for better
predictions because several factors, such as the execution costs,
are dependent on the size of software.


%
\bibliographystyle{abbrv}
\bibliography{cost}
%
%

\balancecolumns

\end{document}